\documentclass[aip,showpacs,superscriptaddress,onecolumn,preprint]{revtex4}%
\usepackage{supertabular}
\usepackage{longtable}
\usepackage{amsfonts}
\usepackage{amsmath}
\usepackage{amssymb}
\usepackage{graphicx}%
\begin{document}
\title{Electrical and optical properties of fluid iron from compressed to expanded regime}
\author{Cong Wang}
\affiliation{Institute of Applied Physics and Computational Mathematics, P.O. Box 8009, Beijing 100088, People's Republic of China} \affiliation{Center for Applied Physics and Technology, Peking University, Beijing 100871, People's Republic of China}
\author{Yujuan Zhang}
\affiliation{Institute of Applied Physics and Computational
Mathematics, P.O. Box 8009, Beijing 100088, People's Republic of
China}
\author{Ze-Qing Wu}
\affiliation{Institute of Applied Physics and Computational Mathematics, P.O. Box 8009, Beijing 100088, People's Republic of China}
\author{Ping Zhang}
\thanks{Corresponding author: zhang\underline{
}ping@iapcm.ac.cn}
\affiliation{Institute of Applied Physics and Computational Mathematics, P.O. Box 8009, Beijing 100088, People's Republic of China} \affiliation{Center for Applied Physics and Technology, Peking University, Beijing 100871,
People's Republic of China}

\begin{abstract}
Using quantum molecular dynamics simulations, we show that the
electrical and optical properties of fluid iron change drastically
from compressed to expanded regime. The simulation results reproduce
the main trends of the electrical resistivity along isochores and
are found to be in good agreement with experimental data. The
transition of expanded fluid iron into a nonmetallic state takes
place close to the density at which the constant volume derivative
of the electrical resistivity on internal energy becomes negative.
The study of the optical conductivity, absorption coefficient, and
Rosseland mean opacity shows that, quantum molecular dynamics
combined with the Kubo-Greenwood formulation provides a powerful
tool to calculate and benchmark the electrical and optical
properties of iron from expanded fluid to warm dense region.
\end{abstract}


\maketitle

\setcounter{MaxMatrixCols}{10}

\bigskip

\section{INTRODUCTION}

Theoretical modelings and direct experimental measurements of
matters under extreme conditions are full of challenges and crucial
interests in astrophysics, inertial confinement fusion, as well as
materials science \cite{Martorell2013,Lindl2004,Tateno2010,Dai2012}.
Due to its technological, geological, and sociological importance,
iron is one of the most studied materials. Considering liquid iron
changes gradually from a compressed into a expanded regime until the
density is lowered by a factor of 3 to 5, a continuous transition
from the condensed state into a gaseous state can be observed
\cite{Korobenko2012,Clerouin2008}. In this procedure, the expanded
metal may lose its metallic properties and change into a nonmetallic
fluid. Other thermal properties, i.e., the equation of state,
radical distribution function, and optical properties may also vary
a lot.

Recently, the development in experimental technique has made it
possible to measure the equation of states and electrical
resistivity at temperatures of the order of 10000 K
\cite{Korobenko2007}. For instance, the wide range of the
liquid-vapor phase transition was investigated for mercury and
alkali metals (Cs and Rb) experimentally
\cite{Franz1980,Kikoin1967}. However, these experiments did not find
any discontinuities in thermodynamic functions except for those
found in the liquid-vapor phase transition region. It should be
noted that recent experiments have performed direct measurements on
the electrical resistivity and provided some evidence for the
existence of a first-order metal-nonmetal (MNM) transition in liquid
aluminum \cite{Korobenko2007,Clerouin2008}. Then, Korobenko and
Rakhel have performed measurements of electrical resistivity and
caloric equation of state for fluid iron to investigate the MNM
transition induced by thermal expansion
\cite{Korobenko2012,Korobenko2011}. In spite of the dramatic
progress, the data reported therein presented a rather narrow range
for the internal energy and pressure, and precluded from a precise
determination of the characteristic density, where the constant
volume derivation changes its sign.

To study the complex behavior of fluid iron from compressed to
expanded regime, we have performed quantum molecular dynamics (QMD)
simulations in a wide range of density and temperature, including
the whole domain where the experimental data were obtained. Here, we
show that the QMD simulations combined with the Kubo-Greenwood
formulation provide a powerful tool to predict the electrical and
optical properties of iron from expanded fluid to warm dense region.

\section{COMPUTATIONAL METHODS}

The basic quantum mechanical density functional theory (DFT) forms
the basis of our calculations, and detailed simulations have been
performed by using Vienna \emph{ab initio} Simulation Package (VASP)
\cite{Kresse1993,Kresse1996}. In our calculations, electrons are
treated fully quantum mechanically by employing a plane-wave finite
temperature DFT within the local spin density approximation (LSDA).
The electronic wave functions are calculated with the projector
augmented wave framework \cite{Blochl1994}. To evolve the ion
trajectories, molecular dynamic simulations have been performed in
the isokinetic ensemble in full thermodynamic equilibrium, with
equal electron and ion temperatures. The electron temperature has
been kept constant according to the Fermi-Dirac distribution, and
ion temperature is controlled by No\'{s}e thermostat
\cite{Hunenberger2005,Evans1985}. All of the dynamic simulations
were lasted 20000 steps with time steps of 2 $\sim$ 4 fs with
respect to different densities and temperatures. In each step, 32
$\sim$ 64 ions are included in a supercell with periodic boundary
conditions, ion positions are then updated classically according to
the electrostatic forces determined within Kohn-Sham construction.
The densities vary from 1.57 g/cm$^{3}$ to 23.55 g/cm$^{3}$
($V/V_{0}$ = 5 $\sim$ $1/3$) with temperatures of 4000 to 20000 K.
The convergence of the calculations has been checked, and a
plane-wave cutoff energy of 600 eV is employed. Brillouin zone
sampling was performed at the mean value point
\cite{Baldereschi1973}.

After equilibrium, ten snapshots have been randomly selected to
perform electronic structure calculations, where a $4\times4\times4$
Monkhorst-Pack $k$-point mesh is used. Then, the dynamical
conductivity $\sigma(\omega)=\sigma_{1}(\omega)+i\sigma_{2}(\omega)$
is evaluated through Kubo-Greenwood formula as averages of the
selected configurations. The dc conductivity ($\sigma_{dc}$) follows
from the static limit $\omega\rightarrow0$ of $\sigma_{1}(\omega)$.
In the Chester-Thellung version \cite{Chester1961}, the kinetic
coefficients $\mathcal{L}_{ij}$ are described as
\begin{equation}
\mathcal{L}_{ij}=(-1)^{i+j}\int
d\epsilon\hat{\sigma}(\epsilon)(\epsilon
-\mu)^{(i+j-2)}(-\frac{\partial f(\epsilon)}{\partial\epsilon}%
),\label{coefficient}%
\end{equation}
with $f(\epsilon)$ being the Fermi-Dirac distribution function and
$\mu$ the chemical potential. The electronic thermal conductivity
$K_{e}$ is given by
\begin{equation}
K_{e}=\frac{1}{T}(\mathcal{L}_{22}-\frac{\mathcal{L}_{12}^{2}}{\mathcal{L}_{11}%
}).\label{thermal}%
\end{equation}

\section{RESULTS}

\begin{figure}
  \includegraphics[width=7.0cm]{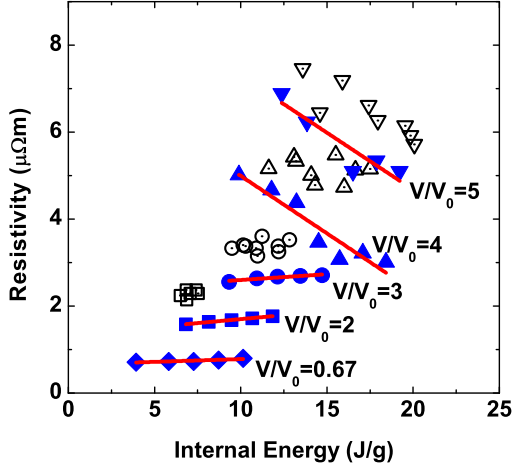}\\
  \caption{(Color online) Electrical resistivity versus internal energy  along five isochores.
  The open symbols represent experimental data, and those
  solid blue ones denote the QMD simulation results
  (diamonds, squares, circles, up triangles, and down triangles correspond
  to $V/V_{0}$=0.67, 2.0, 3.0, 4.0, and 5.0, respectively).
  The red lines are linear fits of the data obtained from QMD results.}\label{fig1}
\end{figure}

Based on QMD simulations, the general trends of the electrical
resistivity as well as the Rosseland mean opacity are concentrated
in this work. It is, as a consequence, interesting to explore not
only to get insight into the interior physical properties of fluid
iron but also to examine a series of theoretical models and
experiments. The electrical resistivity versus internal energy along
the isochores is shown in Fig. \ref{fig1}. The isochores correspond
to the following relative volume values: $V/V_{0}$ = 0.67, 2.0, 3.0,
4.0, and 5.0 ($V_{0}$ = 7.13 cm$^{3}$/mol). In the expanded regime,
the simulated results are in reasonable agreement with those
detected experimentally \cite{Korobenko2012}. Resistivity from QMD
simulations, which is known to underestimate electronic gap by using
LDA \cite{Militzer2003}, is underestimated by about 30\% compared
with experiments. Despite this fact, QMD electrical resistivity
still reproduce the general trend observed experimentally. In the
compressed and some of the expanded region ($V/V_{0} \leq 3$), the
electrical resistivity increases with the internal energy
(temperature) along isochores, and demonstrates a typical behavior
of a metal. However, the resistivity decreases with internal energy
and resembles the behavior of a semiconductor at larger expansions.
The crossover regime locates at the resistivity attaining 3 $\sim$ 5
$\mu\Omega$m, which is very close to the values of
($\sigma_{min}$)$^{-1}$ according to the Mott's criterion
\cite{Mott1984} for the minimum electrical conductivity for metals.

At relative volumes of $V/V_{0}\leq3$, Drude model, which has been
demonstrated to well describe the electrical properties of a simple
metal around normal density, is used to interpret the optical
conductivity spectra in the metallic state for fluid iron. The
optical conductivity or ac conductance obtained from QMD simulations
can be fitted with the following relationship
\begin{equation}\label{drude}
    \sigma_{D}(\omega)=\frac{ze^{2}\tau/m_{e}}{1+\omega^{2}\tau^{2}},
\end{equation}
where $z$ denotes the conducting electron number density, $e$ is the
electronic charge, $\tau$ is the relaxation time, and $m_{e}$ is the
electron mass.

At lower densities ($V/V_{0}\geq4$), the resistivity, characterized
by a much larger value in the magnitude than the metallic values,
decreases with temperature. In this regime, the central peak of the
spectrum for the optical conductivity is shifted, and the
Drude-Smith model \cite{Smith2001}, which introduces the memory
effects of the successive collisions, is more appropriate to fit the
QMD results instead of the Drude formula. The one collision time
formulation for Drude-Smith model reads
\begin{equation}\label{drude-smith}
    \sigma_{DS}(\omega)=\frac{n^{\ast}e^{2}\tau/m_{e}}{1+\omega^{2}\tau^{2}}[1+\frac{c(1-\omega^{2}\tau^{2})}{1+\omega^{2}\tau^{2}}],
\end{equation}
where $c$ (between 0 and $-$1) is the parameter of the memory
effect. $c$ is negative when the central peak of the optical
conductivity is shifted to a positive-energy value. The values
$c$=$-$1 and $c$=0 correspond to the vanishing of dc conductivity
and the classical Drude model, respectively.

\begin{figure}
  \includegraphics[height=7.0cm]{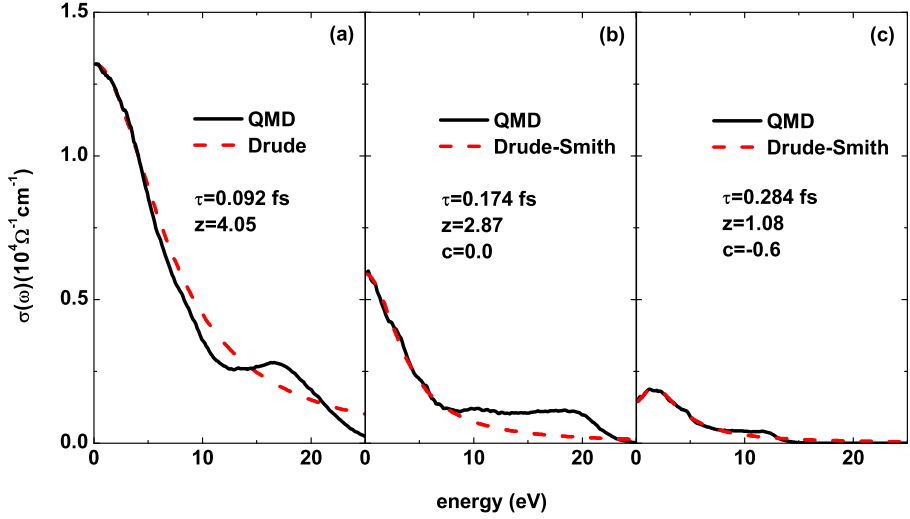}\\
  \caption{(Color online) Optical conductivities and their fits at 10000 K for three
relative volume values: (a) $V/V_{0}$=0.67, (b) $V/V_{0}$=2.0, and
(c) $V/V_{0}$=5.0.}\label{fig2}
\end{figure}

Here, the relaxation time $\tau$, the number density of the
conducting electrons $z$, and parameter $c$ have been adjusted to
get the best fit for the optical conductivity spectrum from QMD
simulations in the low energy range $\hbar\omega\leq5$ eV. An
example of the fit by using Drude formula is shown in Fig.
\ref{fig2} (a) for the compressed fluid with the corresponding
parameters of $V/V_{0}$ = 0.67 and $T$ = 10000 K. A good fit is
obtained for the relaxation time $\tau$ = 0.092 fs and the number of
the conducting electrons per ion $z$ = 4.05. In this case, iron is
still a good metal. In the expanded region, the central peak is
shifted to about 0.1 eV at $V/V_{0}$ = 2 [see Fig. \ref{fig2} (b)],
and a good fit is obtained for $z$ = 2.87, $\tau$ = 0.174 fs, $c$ =
0.0. At larger expansions, the shift of the central peak and the
depression of the dc conductivity become more pronounced. One can
find in Fig. \ref{fig2} (c) for $V/V_{0}$ = 5 that the number of
conducting electrons per ion (in the Drude description) is reduced
to 1.08, and the central peak is shifted to 1.6 eV.

Due to the concurrence of the density and temperature effects,
thermal-physical properties of fluid metals from compressed to
expanded regime are difficult to be treated theoretically. As a
consequence, the MNM transition phenomenon in fluid iron is still a
fundamental and long-standing issue. Well accepted values (3000
$\sim$ 5000 $\Omega^{-1}$cm$^{-1}$) of the minimum dc conductivity
for typical metals were determined based on the experimental
observations. In this work, we are able to determine this value for
fluid iron through the simulated optical conductivity spectrum. The
optical conductivity along the isochore $V/V_{0}\leq2$ can be well
fitted by the Drude formula, and therefore corresponding to a good
metal. At larger expansions ($V/V_{0}\geq3$), the central peak of
the optical conductivity spectrum shifts to positive energy values
with a remarkable part of the conducting electrons reduced, and
nonmetallic states are observed. Thus, the minimum metallic
conductivity for fluid iron is higher than 4000
$\Omega^{-1}$cm$^{-1}$ (corresponding to $V/V_{0}=3$) but it is
lower than the dc conductivity at $V/V_{0}=2$, i.e., 6000
$\Omega^{-1}$cm$^{-1}$. We should note that, it is not enough to
justify a metallic state merely through the indication of the
negative slope, because the simulation results indicate a negative
value of $(\frac{\partial\sigma_{dc}}{\partial T})_{V}$ even for the
isochore $V/V_{0}=3$.

\begin{figure}
  \includegraphics[height=7.0cm]{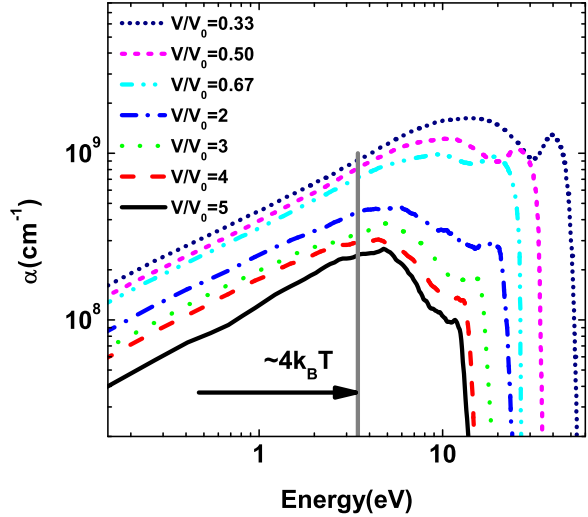}\\
  \caption{(Color online) Absorption coefficient $\alpha(\omega)$ as a function of photon energy at
  T=10000 K for $V/V_{0}$=5.0 (solid line), 4.0 (dashed line), 3.0 (dotted line), 2.0 (dashed-dotted line),
  0.67 (dashed-dotted-dotted line), 0.50 (short dashed line), and 0.33 (short dotted line). Bar at $4k_{B}T$ represents maximum region of contribution to the Rosseland mean opacity.}\label{fig3}
\end{figure}

We also investigate the trends in the optical absorption
coefficient, which is shown in Fig. \ref{fig3} along the 10000 K
isotherm. Rapid increase in the absorption coefficient has been
observed in the photon energy range of a few eV. Let us recall here
that in a standard opacity calculation, the free-free contribution
(also called inverse bremsstrahlung), which stems from the photon
scattering off free electrons belonged to the ionic Coulomb field,
governs this feature in the low photon energy region. A classical
approximation, known as the Kramers formula, suggests
$\sigma_{1}(\omega)\sim1/\omega^{3}$ at low densities. In QMD
calculations, although the Kubo-Greenwood formula does not
explicitly distinguish each contribution to the total absorption
coefficient as in a standard opacity calculation, some phenomena
clearly indicate that the free-free contribution still works in the
low photon energy range. This can be substantiated by first noting
from Fig. \ref{fig3} that the magnitude of the free-free
contribution is proportional to $z$. In the low photon energy
regime, the absorption coefficients have similar behaviors and vary
by an order of magnitude as $z$ varies from 1.0 to 4.0. Second, high
lying excited states contribute to QMD optical conductivity and
absorption coefficient, so that the power law suggested by Kramers
formula is not followed by QMD calculations. It is important to
stress here that, the power law was built on a classical description
of the electron-ion interaction, where an isolated system contains a
photon, an electron, and a charged ion without screening from the
surrounding media. On the contrary, the Kubo-Greenwood formula used
in the present work introduces a one-electron approximation to
obtain the optical conductivity within the linear response theory
\cite{Harrison1970,Callaway1974}. The scattering cross sections for
an isolated system are included by Kubo-Greenwood formulation
\cite{Perrot1996,Iglesias1996,Crowley2001}, and the Kramers formula
for the free-free contribution can be superceded by the current QMD
simulations.

\begin{figure}
  \includegraphics[height=7.0cm]{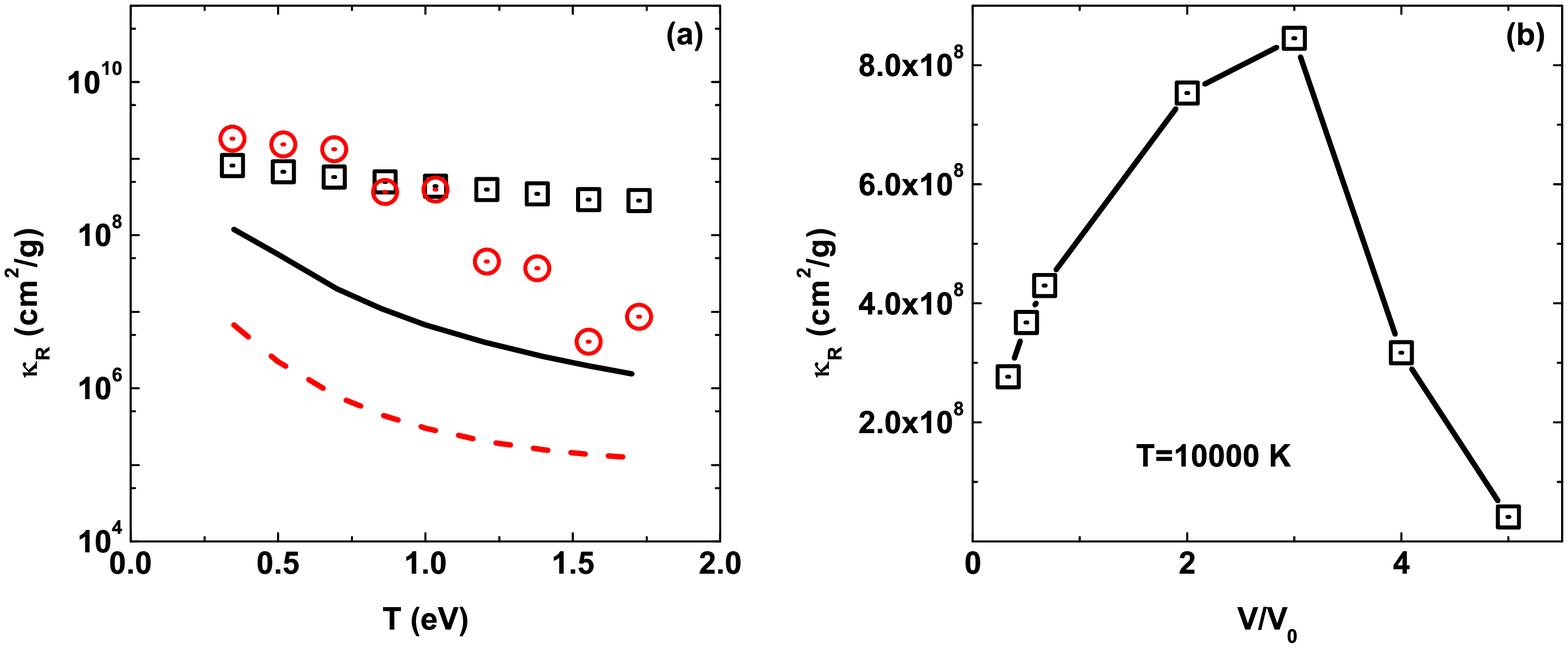}\\
  \caption{(Color online) Variation of the QMD and DCA Rosseland mean opacities:
  (a) as a function of temperature and for fixed densities of $V/V_{0}$=0.67 (QMD: open squares,
  DCA: solid line) and $V/V_{0}$=4.0 (QMD: open circles, DCA: dashed line);
  (b) as a function of relative volume and for a fixed temperature of $T$=10000 K (QMD: open squares).}\label{fig4}
\end{figure}

A direct comparison of the Rosseland mean opacities obtained from
QMD and from a detailed configuration accounting (DCA) model
\cite{Yan2002} has been shown in Fig. \ref{fig4}(a). In DCA model,
the absorption coefficients are calculated with considering the
appropriate isolated atom photoionization (bound-free), inverse
bremsstrahlung (free-free), scattering cross sections, and
bound-bound transition arrays. In Fig. \ref{fig4}(a), we show a
comparison of the QMD and DCA opacities at densities of $V/V_{0}=$
0.67 and 4.0 by increasing temperature. As indicated in Fig. 4(a),
disagreement exists in the whole thermodynamic regime we explore.
The DCA Rosseland mean opacity, which is dominated by the free-free
contribution at temperatures of a few eV, decreases with temperature
and increases with density. However, in QMD simulations, a crossover
region of the mean opacities around 10000 K is observed for the two
densities. To get insight into this particular behavior, we show in
Fig. \ref{fig4}(b) the variation of the Rosseland mean opacity for a
fixed temperature of $T=10000$ K, and as the relative volume varies
from 0.33 to 5. At $V/V_{0} \leq 2$, the optical conductivity, as
well as the absorption coefficient behaves as a simple metal, thus
the Rosseland mean opacity is proportional to $\alpha(\omega)$
around $4k_{B}T$ (see Fig. \ref{fig3}). As a result, the increase of
the Rosseland mean opacity with relative volume is observed. Beyond
the relative volume $V/V_{0} =$ 3, QMD results show a continuous
decrease, which indicates a gradual decrease of the charge carriers
in the nonmetal regime. In contrast, DCA calculations do not reveal
this crossover region, and the difference can be traced back to the
limitation in determining the continuum lowering of the ionization
energy and the pressure ionization model used in DCA code.

\section{SUMMARY}

In summary, we have performed QMD simulations for fluid iron from
compressed to expanded region. The calculated resistivity versus
internal energy along isochores are in good agreement with the
experimental data. The observation of the negative slope of the
dependence of the resistivity on internal energy, combined with the
shift of the central peak in the optical conductivity, indicates
transition of fluid iron into a nonmetallic state. The optical
properties of the system are analyzed through the computation of the
absorption coefficient and the Rosseland mean opacity, which reveal
the continuous decrease in the charge carriers as the system looses
its metallic properties. This analysis confirms that, the QMD
simulations are powerful tools to validate classical models by
providing a consistent set of electrical and optical properties from
the same simulation.

\section{ACKNOWLEDGMENTS}

This work was supported by NSFC under Grants No. 11275032, No.
11005012 and No. 51071032, by the foundation for Development
of Science and Technology of China Academy of Engineering Physics
under Grant No. 2013B0102019, by the National Basic Security Research
Program of China, and by the National High-Tech ICF
Committee of China.

\end{document}